# PARISROC, a Photomultiplier Array Integrated Read Out Chip

S. Conforti Di Lorenzo[a], J.E. Campagne[b], F. Dulucq[a], C. de La Taille[a], G. Martin-Chassard[a],
M. El Berni[a], W. Wei[c]

[a] OMEGA/LAL/IN2P3, centre universitaire BP34 91898 ORSAY Cedex, France
[b] LAL/IN2P3, centre universitaire BP34 91898 ORSAY Cedex, France
[c] IHEP, Beijing, China

conforti@lal.in2p3.fr

## Abstract

PARISROC is a complete read out chip, in AMS SiGe 0.35 µm technology [1], for photomultipliers array. It allows triggerless acquisition for next generation neutrino experiments and it belongs to an R&D program funded by the French national agency for research (ANR) called PMm2: "Innovative electronics for photodetectors array used in High Energy Physics and Astroparticles" [2] (ref.ANR-06-BLAN-0186). The ASIC integrates 16 independent and auto triggered channels with variable gain and provides charge and time measurement by a Wilkinson ADC and a 24-bit Counter. The charge measurement should be performed from 1 up to 300 photo-electrons (p.e.) with a good linearity. The time measurement allowed to a coarse time with a 24-bit counter at 10 MHz and a fine time on a 100ns ramp to achieve a resolution of 1 ns. The ASIC sends out only the relevant data through network cables to the central data storage. This paper describes the front-end electronics ASIC called PARISROC.

## I. Introduction

The PMm$^2$ project proposes to segment the large surface of photodetection [3] in macro pixel consisting of an array (2*2m) of 16 photomultipliers connected to an autonomous front-end electronics (Figure 1) and powered by a common High Voltage. These large detectors are used in next generation proton decay and neutrino experiment i.e. the post-SuperKamiokande detector as those that will take place in megaton size water Cerenkov or 100kt size liquid scintillator one. These news detectors will require very large surfaces of photo detection at a moderate cost. This R&D [2] involves three French laboratories (LAL Orsay, LAPP Annecy, IPN Orsay) and ULB Brussels for the DAQ.

LAL Orsay is in charge of the design and tests of the readout chip named PARISROC which stands for Photomultiplier ARrray Integrated in Si-Ge Read Out Chip.

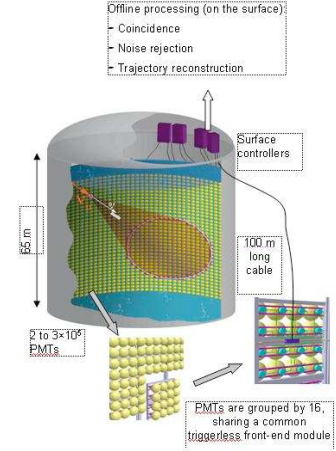

Figure 1: Principal of PMm2 proposal for megaton scale Cerenkov water tank.

## II. PARISROC architecure.

### A. Global architecture

The ASIC PARISROC (Figure 2) is composed of 16 analog channels managed by a common digital part.

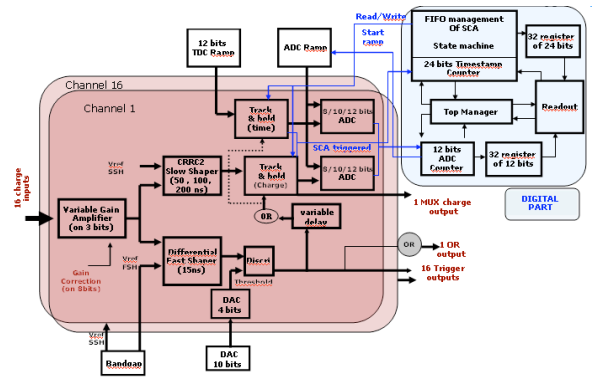

Figure 2: PARISROC global schematic.

Each analog channel (Figure 3) is made of a voltage preamplifier with variable and adjustable gain. The variable gain is common for all channels and it can change thanks to the input variable capacitance on 3 bits. The gain is also tuneable channel by channel to adjust the input PMTs gain non homogeneity, thanks to the switched feedback capacitance on 8 bits.

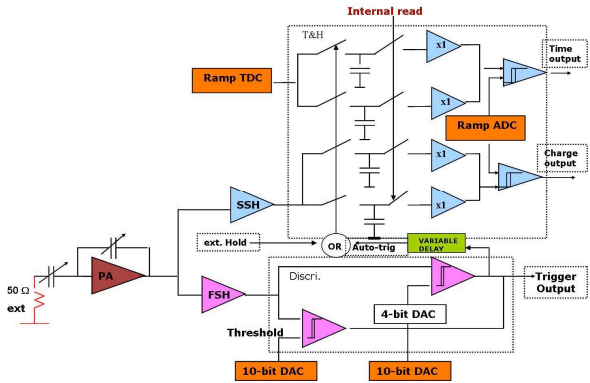

Figure 3: One channel schematic.

The preamplifier is followed by a slow channel for the charge measurement in parallel with a fast channel for the trigger output. The slow channel is made by a slow shaper followed by an analog memory with a depth of 2 to provide a linear charge measurement up to 50 pC; this charge is converted by a Wilkinson ADC (8,9 or 12 bits). One follower OTA is added to deliver an analog multiplexed charge measurement. The fast channel consists in a fast shaper followed by 2 low offset discriminators to auto-trig down to 50 fC. The thresholds are loaded by 2 internal 10-bit DACs common for the 16 channels and an individual 4-bit DAC for one discriminator. The 2 discriminator outputs are multiplexed to provide only 16 trigger outputs. Each output trigger is latched to hold the state of the response until the end of the clock cycle. It is also delayed to open the hold switch at the maximum of the slow shaper. An "OR" of the 16 trigger gives a 17th output. For each channel, a fine time measurement is made by an analog memory with depth of 2 which samples a 12-bit TDC ramp of 100 ns, common for all channels, at the same time of the charge. This time is then converted by the Wilkinson ADC. The two ADC discriminators have a common ramp, of 8/10/12 bits, as threshold to convert the charge and the fine time. In addition a bandgap bloc provides all voltage references.

### B. Digital part.

On overview of the digital part is given in figure 4. The digital bloc manages the track and hold system like a FIFO and starts and stops all the counters [4]. All the data are serialized to be sent out.

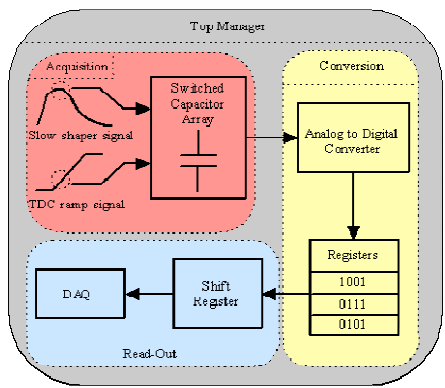

Figure 4: Digital part overview.

There are two clocks: one at 40 MHz for the analog to digital conversion and for the track and hold management, the second at 10MHz for timestamp and readout.

The readout format is 52 bits: 4 bits for channel number + 24 bits for timestamp + 12 bits for charge conversion + 12 bits for fine time conversion. The readout is selective: only the hit channels are read; so the maximum readout time will be 100µs if all channels are hit.

### III. MEASUREMENTS AND SIMULATION.

### A. General tests.

A dedicated test board has been designed and realized for testing the ASIC (Figure 5). Its aim is to allow the characterization of the chip and the communication between photomultipliers and ASIC. This is possible thanks to a dedicated Labview program that allows sending the ASIC configuration (slow control parameters, ASIC parameters, etc) and receiving the output bits via an USB cable connected to the test board. The Labview is developed by the LAL "Tests group".

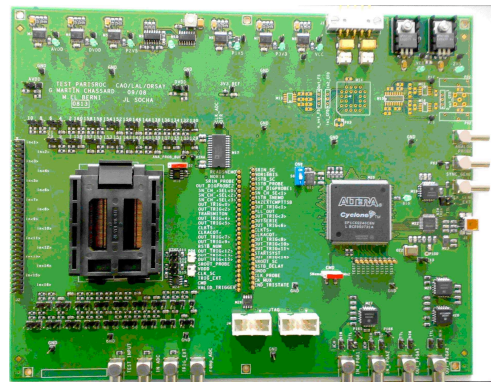

Figure 5: Test board.

*1) Input signal.*

A signal generator is used to create the input charge injected in the ASIC. The signal injected is similar, as possible, to the PMT signal. In Figure 6 is represented the generator input signal and its characteristics. The input signal, used in measurements and simulation, is a triangle signal with 5 ns rise and fall time and 5 ns of duration. This current signal is sent to an external resistor (50 Ohms) and varies from 0 to 5 mA in order to simulate a PMT charge from 0 to 50 pC which represents 0 to 300 p.e. when the PM gain is $10^6$.

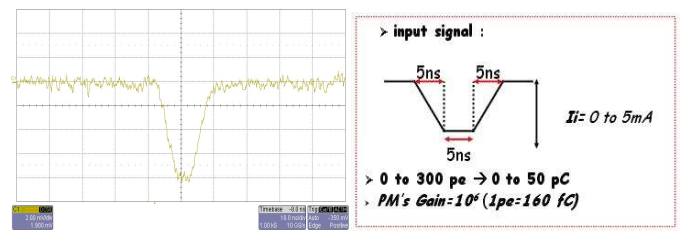

Figure 6: Input signal used for measurements and simulations.

*2) Analog part tests*.

Table 1 lists the simulation and measurements results for the three main blocks of the analog part: Preamplifier, Slow shaper and Fast shaper.

|  | Preamplifier Gain PA=8 Meas. / Sim. | Slow Shaper RC=50ns Meas. / Sim. | Fast Shaper Meas. / Sim. |
| --- | --- | --- | --- |
| Voltage (1 p.e.) | 5mV/5.43mV | 12mV/19mV | 30mV/39mV |
| rms noise/ Noise p.e. | 1mV/468uV 0.2/0.086 | 4mV/2.3mV 0.3/0.125 | 2.5mV/2.4mV 0.08/0.06 |
| (SNR) | 5/12 | 3/8 | 12/16 |

Table 1: Analog part results.

There is a good agreement between measurements and simulation in analog part results except for the noise values. To characterize the noise, the Signal to Noise Ratio (SNR) is calculated with reference to the MIP (1 p.e.). The noise differences are immediately evident: an additional low frequency noise is present in measurement (is now under investigation even if it is supposed to be tied to the power supply noise). A small difference has been noticed in measurement without the USB cable that allowed the communication between the test board and the Labwiev program: an rms noise value of 660µV (0.132 p.e.) for preamplifier and so a SNR value of 8.

Another important characteristic is the linearity. The preamplifier linearity in function of variable feedback capacitor value with an input charge of 10 p.e. and with residuals from -2.5 to 1.35 % is represented on Figure 7. The gain adjustment linearity is good at 2% on 8 bits.

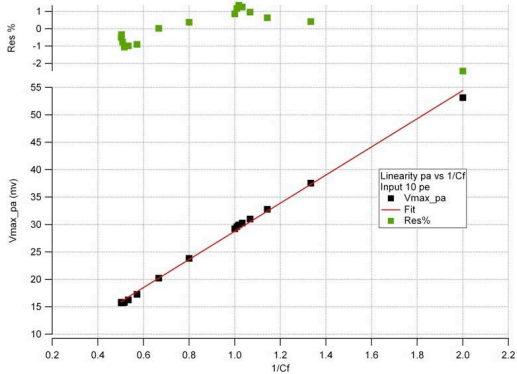

Figure 7: Preamplifier linearity vs feedback capacitor value.

Figure 8 represents the slow shaper linearity for a time constant of 50 ns and a preamplifier gain of 8. The slow shaper output voltage in function of the input injected charge is plotted. Good linearity performances are obtained with residuals better than ± 1%.

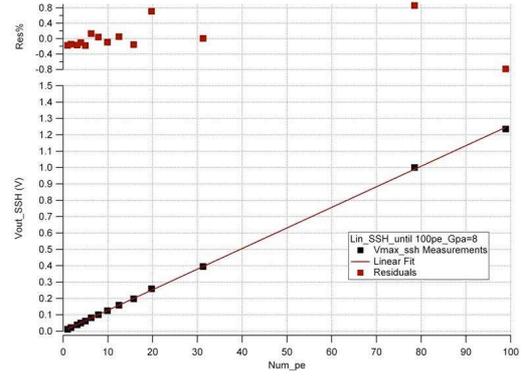

Figure 8: Slow shaper linearity; τ=50 ns and Gpa=8.

In order to investigate the homogeneity among the whole chip, essential for a multichannel ASIC, for the different preamplifier gains is plotted the maximum voltage value for all channels. On Figure 9 is given the gain uniformity. A good dispersion of 0.5%, 1.4% and 1.2% have respectively been obtained for gain 8, 4 and 2. This represents a goal for the ASIC.

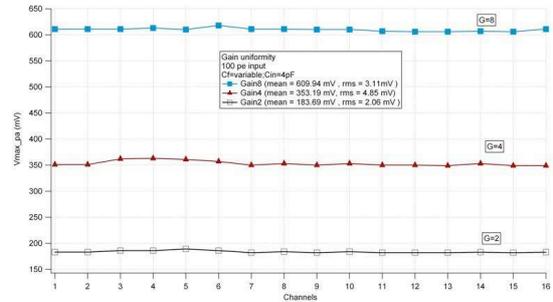

Figure 9: Gain uniformity for Gpa = 8, 4, 2.

### B. DAC Linearity

The DAC linearity has been measured and it consists in measuring the voltage DAC (Vdac) amplitude obtained for different DAC register values. Figure 10 gives the evolution of Vdac as a function of the register for the DAC and the residuals with values from -0.1% to 0.1%.

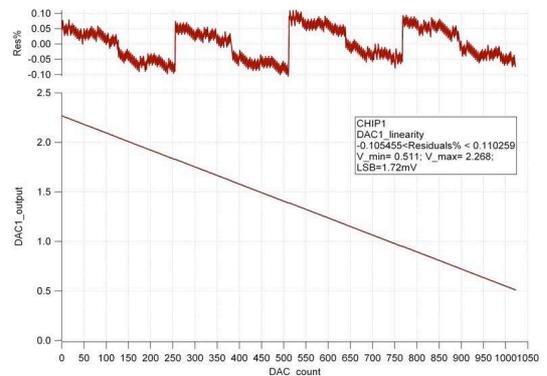

Figure 10:  DAC linearity.

## C. Trigger Output.

The trigger output behaviour was studied scanning the threshold for different injected charges. At first no charge was injected which corresponds to measure the fast shaper pedestal. The result is represented on Figure 11 for each channel. The 16 curves (called s-curves because of their shape) are superimposed that meaning good homogeneity. The spread is of one DAC count (LSB DAC=1.78 mV) equivalent to 0.06 p.e.

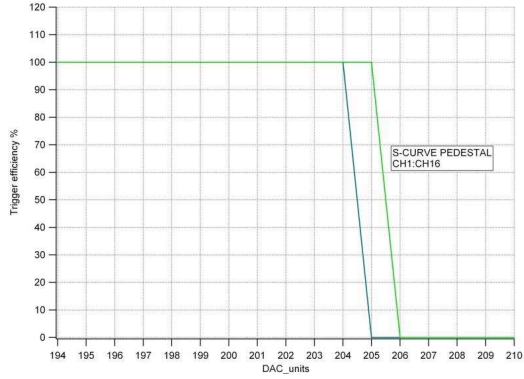

Figure 11: Pedestal S-Curves for channel 1 to 16.

The trigger efficiency was then measured for a fixed injected charge of 10 p.e. On Figure 12 are represented the S-curves obtained with 200 measurements of the trigger for all channels varying the threshold. The homogeneity is proved by a spread of 7 DAC units (0.4 p.e).

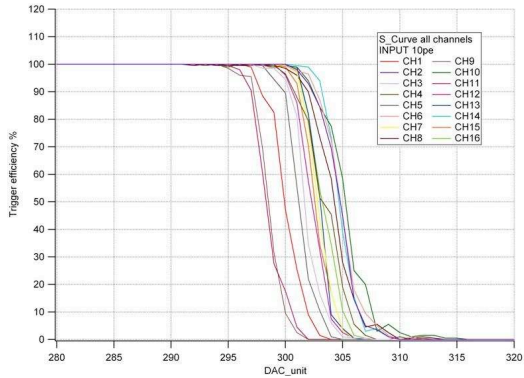

Figure 12: S-Curve for input of 10 p.e. for channel 1 to 16.

The trigger output is studied also by scanning the threshold for a fixed channel and changing the injected charge. Figure 13 shows the trigger efficiency versus the DAC unit with an injected charge from 0 to 300 p.e. and on Figure 14 is plotted the threshold versus the injected charge but only until 0.5 pC.

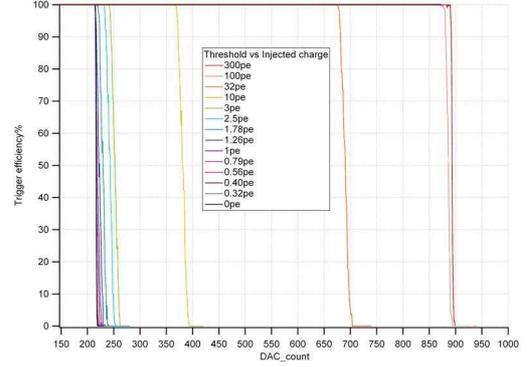

Figure 13: Trigger efficiency vs DAC count up to 300 p.e.

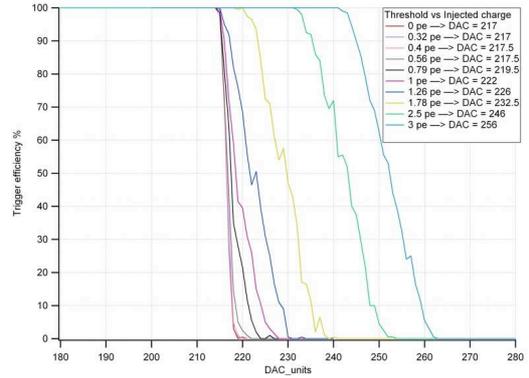

Figure 14: Trigger efficiency vs DAC count until 3 p.e.

In Figure 15 are plotted the 50% trigger efficiency values, extracted from the plot in Figure 14, converted in mV versus the injected charges. A noise of 10fC has been extrapolated. Therefore the threshold is limited to 10 σ noise due to the discriminator coupling.

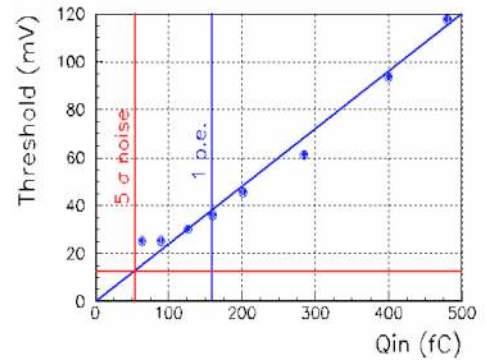

Figure 15: Threshold vs injected charge until 500 fC.

## D. ADC.

The ADC performance has been studied alone and with the whole chain. Injecting to the ADC input directly DC voltages by the internal DAC (in order to have a voltage level as stable as possible) the ADC values for all channels have been measured. The measurement is repeated 10000 times for each channel and in the first panel of the Labview window (Figure 16) the minimal, maximal and mean values, over all acquisitions, for each channel are plotted. In the second panel there is the rms charge value versus channel number with a value in the range [0.5, 1] ADC unit. Finally the third panel shows an example of charge amplitude distribution for a single channel; a spread of 5 ADC counts is obtained.

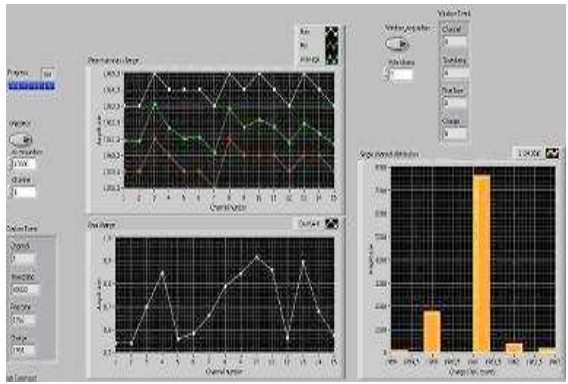

Figure 16: ADC measurements with DC input 1.45V.

The ADC is suited to a multichannel conversion so the uniformity and linearity are studied in order to characterize the ADC behaviour. On Figure 17 is represented the ADC transfer function for the 10-bit ADC versus the input voltage level. All channels are represented and have plots superimposed.

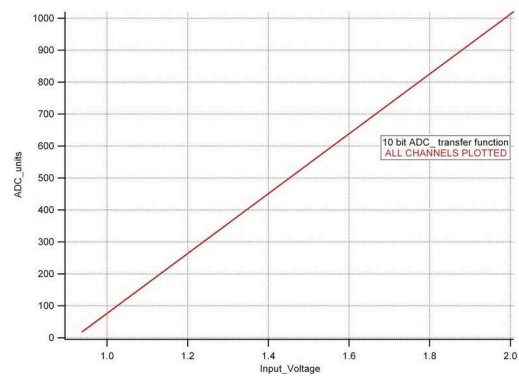

Figure 17: 10-bit ADC transfer function vs input charge.

This plot shows the good ADC uniformity among the 16 channels. In Figure 18 is shown the 12-bit ADC linearity plots with the 25 measurements made at each input voltage level. The average ADC count value is plotted versus the input signal. The residuals from -1.5 to 0.9 ADC units for the 12-bit ADC; from -0.5 to 0.4 for the 10-bit ADC and from -0.5 to 0.5 for the 8-bit ADC prove the good ADC behaviour in terms of Integral non linearity.

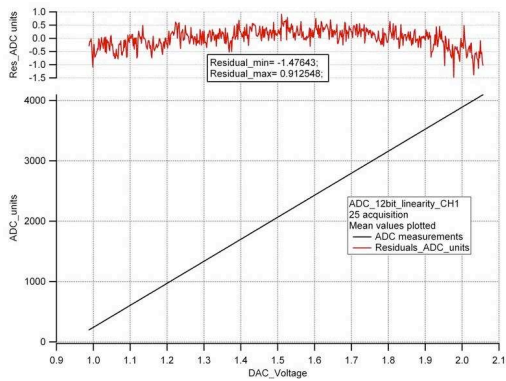

Figure 18: 12-bit ADC linearity.

Once the ADC performances have been tested separately, the measurements are performed on the complete chain. The results of the input signal auto triggered, held in the T&H and converted in the ADC are illustrated in Figure 19 where are plotted the 10-bit ADC counts in function of the variable input charge (up to 50 p.e). A nice linearity of 1.4% and a noise of 6 ADC units are obtained.

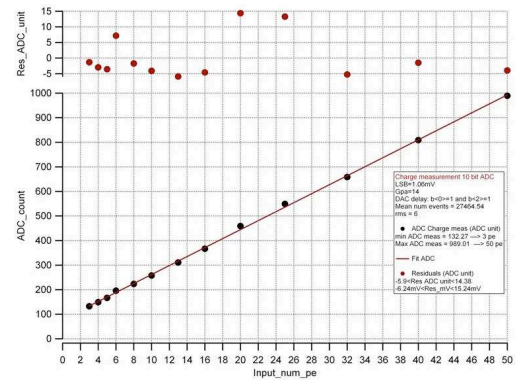

Figure 19: 10-bit ADC linearity.

## IV. CONCLUSION

Good overall performances of the chip PARISROC are obtained: auto trigger signal and digitalization of DATA. Good uniformity and linearity although strange noise performance due to 10 MHz clock noise and a low frequency noise under investigation. A second version of the chip will be submitted in November 09 with an increasing of the dynamic range thanks to 2 preamplifier gains: high gain and low gain; 8/9/10 bits ADC to reduce the p.e. loss below 1% level in case of 5 kHz dark current per PMT and a double fine TAC.

## V. REFERENCES

[1] http://asic.austriamicrosystems.com/
[2] http://pmm2.in2p3.fr/
[3] B. Genolini et al., PMm2: large photomultipliers and innovative electronics for next generation neutrino experiments, NDIP'08 conference. doi:10.1016/j.nima.2009.05.135.
[4] F. Dulucq et al., Digital part of PARISROC: a photomultiplier array readout chip, TWEPP08 conference.